\documentclass[twocolumn,prl,showpacs,preprintnumbers,amsmath,amssymb]{revtex4}
\usepackage{graphicx,epsfig}
\usepackage{dcolumn}
\usepackage{bm}
\begin{document}
\title{Universal power law in the orientational relaxation in thermotropic liquid 
crystals}
\author{Dwaipayan Chakrabarti, Prasanth P. Jose, Suman Chakrabarty, 
and Biman Bagchi} 
\email{bbagchi@sscu.iisc.ernet.in}
\affiliation{Solid State and Structural Chemistry Unit, Indian Institute of 
Science, Bangalore 560012, India}

\date \today

\begin{abstract}

We observe a surprisingly general power law decay at short to intermediate times 
in orientational relaxation in a variety of model systems (both calamitic and 
discotic, and also discrete) for thermotropic liquid crystals. As all these 
systems transit across the isotropic-nematic phase boundary, two power law 
relaxation regimes, separated by a plateau, emerge giving rise to a step-like 
feature (well-known in glassy liquids) in the single-particle second-rank 
orientational time correlation function. In contrast to its probable dynamical 
origin in supercooled liquids, we show that the power law here can originate from 
the thermodynamic fluctuations of the orientational order parameter, driven by the
rapid growth in the second-rank orientational correlation length.
  
\pacs{61.20.Lc,64.70.Md,64.70.Pf}
 
\end{abstract}

\maketitle

%\large

Thermotropic liquid crystals are known to exhibit interesting, often exotic, 
dynamical properties which are subjects of great fundamental and practical 
interests \cite{r1,r2}. For calamitic liquid crystals that comprise rod-like 
molecules, the approach to increasingly ordered mesophases on lowering temperature
involves an isotropic to nematic (I-N) transition and then a nematic to smectic-A 
(N-A) transition that are both believed to be only weakly first order with 
considerable characteristics of continuous transitions \cite{r1,r2,r3,r4,r5,r6}. 
On the other hand, discotic liquid crystals that consist of disc-like molecules 
exhibit a transition from an isotropic to a nematic-discotic phase, and onward to 
a columnar phase rather than a smectic phase \cite{r1,r2}. 
 
Surprisingly, dynamics across the I-N phase transition appears to have been 
investigated mostly at low frequencies or long times (milliseconds to nanoseconds) 
\cite{r1}. However, optical Kerr effect (OKE) measurements \cite{r7} by 
Fayer {\it et al.} have recently revealed a power law relaxation near the 
I-N phase boundary as well as in the nematic phase of rod-like molecules having 
aspect ratios in between 3 and 4 \cite{r8,r9}. While the power law decay appears 
at short (a few picoseconds) to intermediate (a few hundred nanoseconds) times in 
the isotropic phase near the I-N transition \cite{r8}, multiple power laws persist
even at long times in the nematic phase \cite{r9}. Although a few theoretical 
analyses exist in the literature \cite{r8,r10}, the origin and the scope of 
this dominant power law relaxation are poorly understood at present.

In order to understand this power law decay, we have carried out a long and 
extensive molecular dynamics (MD) simulation study of orientational relaxation 
in a variety of model systems for thermotropic liquid crystals, both continuous 
and discrete. The continuous systems in our study consist of {\it ellipsoids of 
revolution}. We have used the Gay-Berne pair potential \cite{r11}, which is an 
elegant generalization of the extensively used isotropic Lennard-Jones potential 
to explicitly incorporate anisotropy in both the attractive and the repulsive 
parts of the interaction with a single-site representation for each ellipsoid of 
revolution \cite{r12}. For a representative calamitic system, we have considered a
system of $576$ {\it prolate} ellipsoids of revolution with an aspect ratio of $3$
that is comparable to those of the rod-like molecules recently studied 
experimentally by Fayer and coworkers. We have investigated the system along an 
isotherm at several densities. 

In the quest of a universal power law behavior, if at all there is any, we have in 
addition undertaken simulations of $500$ {\it oblate} ellipsoids of revolution 
with an aspect ratio of $0.345$. We have used in this case a Gay-Berne pair 
potential that has been modified for disc-like molecules by Bates and Luckhurst 
\cite{r13}. The discotic system has been studied here along an isobar at several 
temperatures.    

The well-known Lebwohl-Lasher (LL) model is a prototype of lattice models 
\cite{r14}, where the particles are assumed to have uniaxial symmetry and 
represented by three-dimensional spins, located at the sites of a simple cubic 
lattice, interacting through pair potential of the form 
$U_{ij} = -\epsilon_{ij}P_{2}(cos \theta_{ij})$. Here $\epsilon_{ij}$ is a 
positive constant $\epsilon$ for nearest neighbor spins $i$ and $j$ and zero 
otherwise, $P_{2}$ is the second rank Legendre polynomial and $\theta_{ij}$ is the
angle between the spins $i$ and $j$. In this work, we have also considered a 
1000-particle LL lattice system to study pure orientational dynamics across the 
I-N transition. The simplicity of the model allows us to study a larger system 
size. In our MD simulations, the system undergoes a transition from the isotropic 
to nematic phase at the temperature $T \simeq0.14$.

\begin{figure}
\epsfig{file=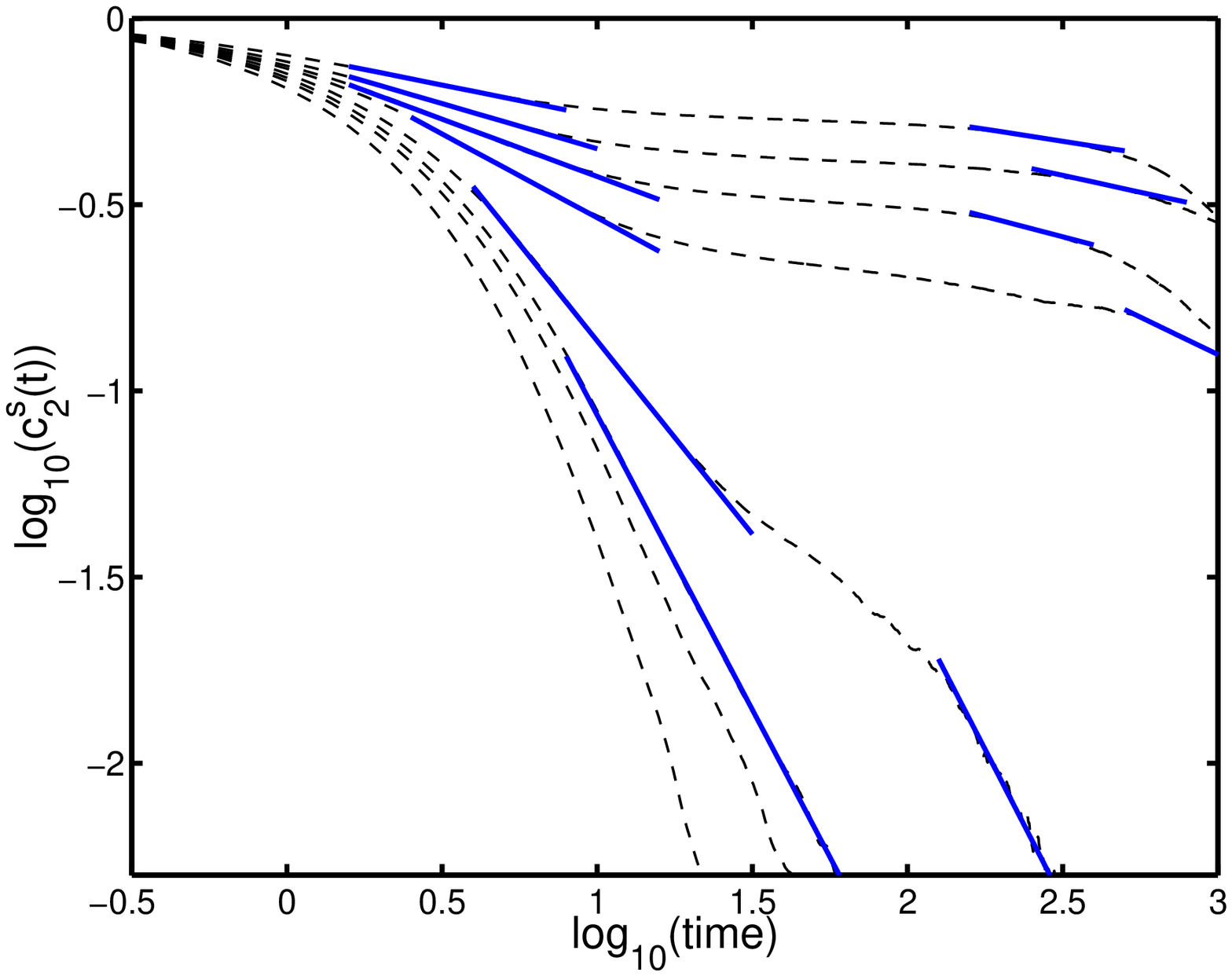,width=7.5cm}
\epsfig{file=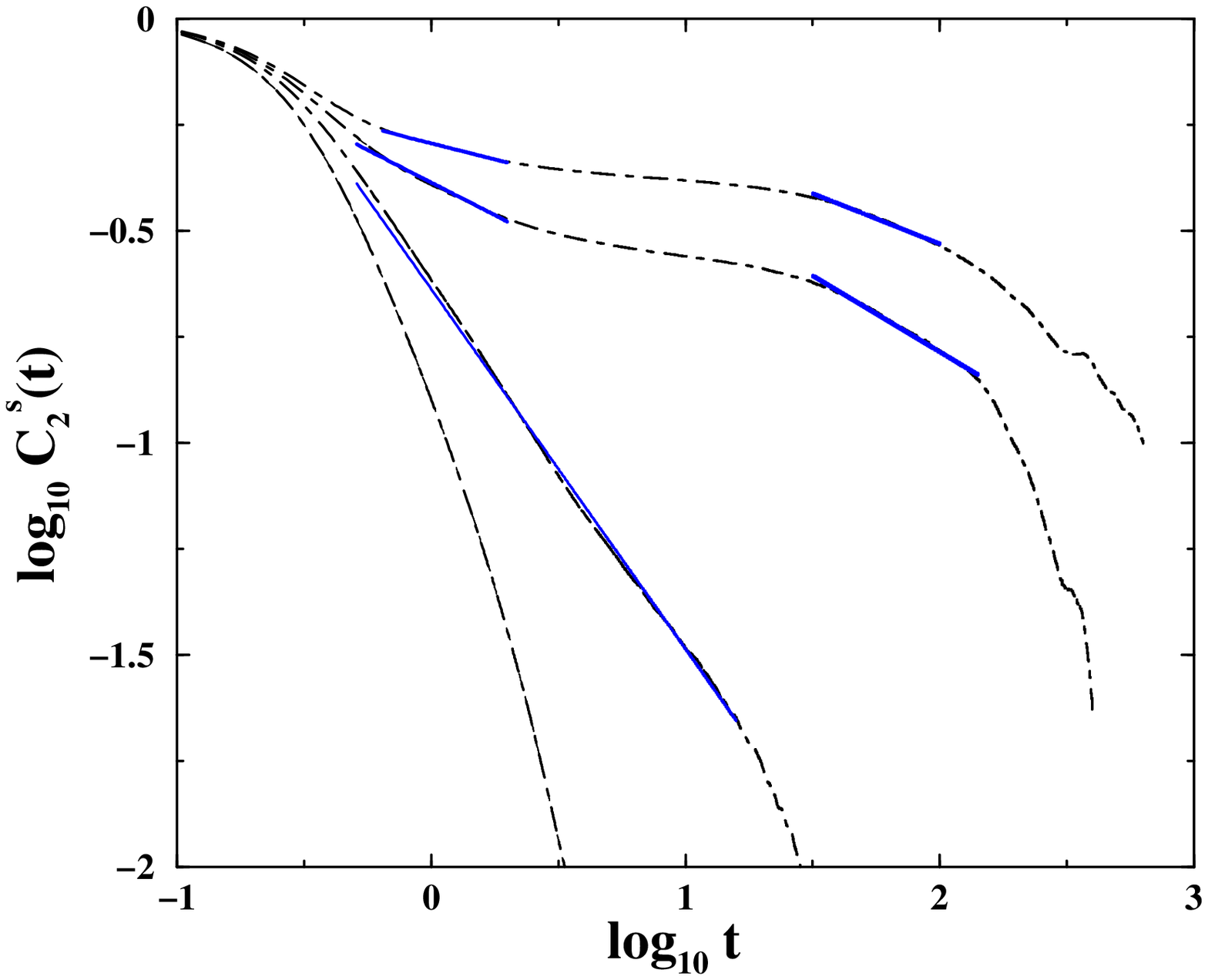,width=7.5cm}
\epsfig{file=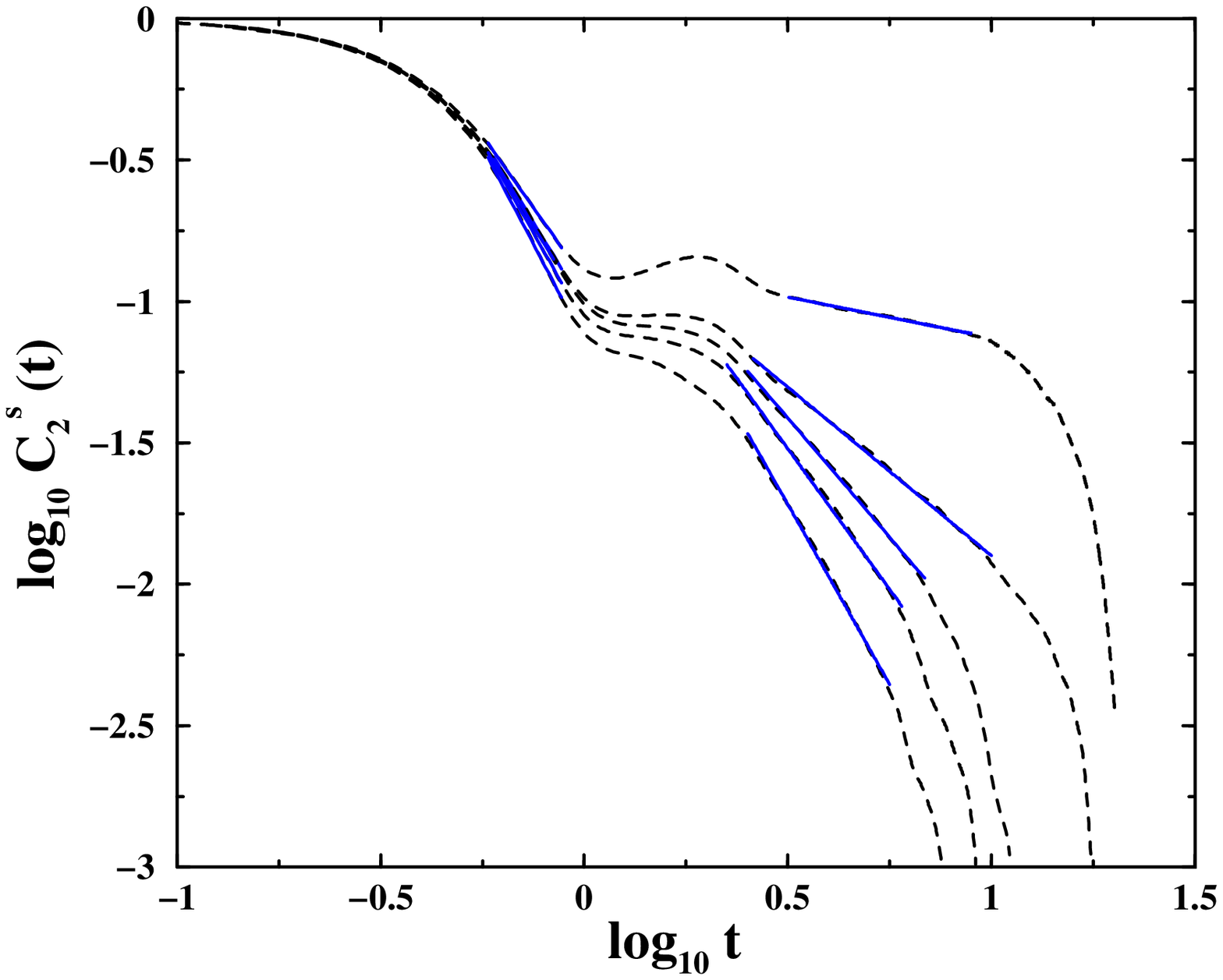,width=7.5cm}
\caption{Time evolution of the single-particle second rank OTCF in log-log plots 
for the (a) calamitic system, (b) discotic system, and (c) lattice system. The 
black lines are simulation data corresponding to increasing order parameter from 
bottom to top. The blue solid lines show the fits to the data over the time 
regimes where the decay follows a power law. (a) Along an isotherm at $T = 1.0$ at 
several densities: 
$\rho = 0.285, 0.295, 0.3, 0.305, 0.31, 0.315, 0.32$, and $0.33$. The I-N 
transition density is brackted by $\rho = 0.305$ and $\rho = 0.315$ with 
$\rho = 0.31$ falling on the transition region. (b) Along an isobar at $P = 25.0$ 
at several temperatures: $ T = 2.991, 2.693, 2.646$, and $2.594$. The I-N 
transition temperature is bracketed by $T=2.693$ and $T = 2.646$. (c) At several 
temperatures: $T = 1.213, 1.176, 1.160, 1.149, 1.134$.} 
\end{figure}
In Figs. 1(a), 1(b), and 1(c), we show in log-log plots the time evolution of the 
single-particle, second rank orientational time correlation function (OTCF) 
observed for the calamitic, discotic and lattice systems, respectively, across the
I-N phase transition \cite{r15}. The I-N transition is marked by a jump in the 
orientational order parameter S, defined for an N-particle system as the largest 
eigenvalue of the ordering matrix Q: $Q_{\alpha\beta} = \frac{1}{N}\sum_{i=1}^{N}
\frac{1}{2}(3e_{i\alpha}e_{i\beta}-\delta_{\alpha\beta})$, where $e_{i\alpha}$ is 
the $\alpha$-component of the unit orientation vector $\hat{\bf e}_{i}$ along the
principal symmetry axis of the $i$-th ellipsoid of revolution in the space fixed 
frame \cite{r16}. Thy single-particle second rank OTCF $C_{2}^{s}(t)$ is defined 
by
\begin{equation}
C_{2}^{s}(t)={\frac{\langle{\displaystyle \sum_{i} P_{2}({\bf \hat e}_{i}(0)\cdot 
{\bf \hat e}_{i}(t))}\rangle}{
\langle{\displaystyle \sum_i P_{2}({\bf \hat e}_{i}(0)\cdot {\bf \hat e}_{i}(0))}
\rangle}},
\label{cs2}
\end{equation}
where the angular brackets stand for statistical averaging.Note in these figures 
{\it the emergence of the power law decay} at short to intermediate times near the 
I-N phase boundary. Figures 1(a), 1(b), and 1(c) further show that as the systems 
transit across the I-N phase boundary, two power law relaxation regimes, separated
by a plateau, appear giving rise to a step-like feature. {\it The intriguing 
feature is the universality in qualitative behavior given that widely different 
model systems are under consideration.} Although the values of the power law
exponents vary considerably from one system to another, certain trends are 
apparent. The power law exponent values decrease monotonically for all the systems
on approaching the I-N phase boundary from the isotropic side and undergo a rather
sharp drop to a value below $0.45$ on crossing the boundary. The only exception to
this is the early power law exponent for the LL lattice system where the two power
law relaxation regimes are evident even in the isotropic phase with a rather weak 
temperature sensitivity for the early power law exponent. In this case, a hump 
appears in the plateau region on approaching the I-N phase boundary. This
can be attributed to the inertial effects because of low damping of the
orientational motion. Near the I-N phase boundary where the continuous systems
show a single power law relaxation regime, the exponent has a value above 
($1.582$) or below ($0.848$) unity depending upon whether system is calamitic or 
discotic. The values of the exponents in these systems seem to rule out grouping 
them together in a common universality class. The weakly first order nature
of the I-N transition also precludes the existence of such strict universality. 
Note that the overall relaxation behavior looks surprisingly similar to the decay 
of self-intermediate scattering function in supercooled liquids where such power 
laws are well-known \cite{r17}.

\begin{figure}
\epsfig{file=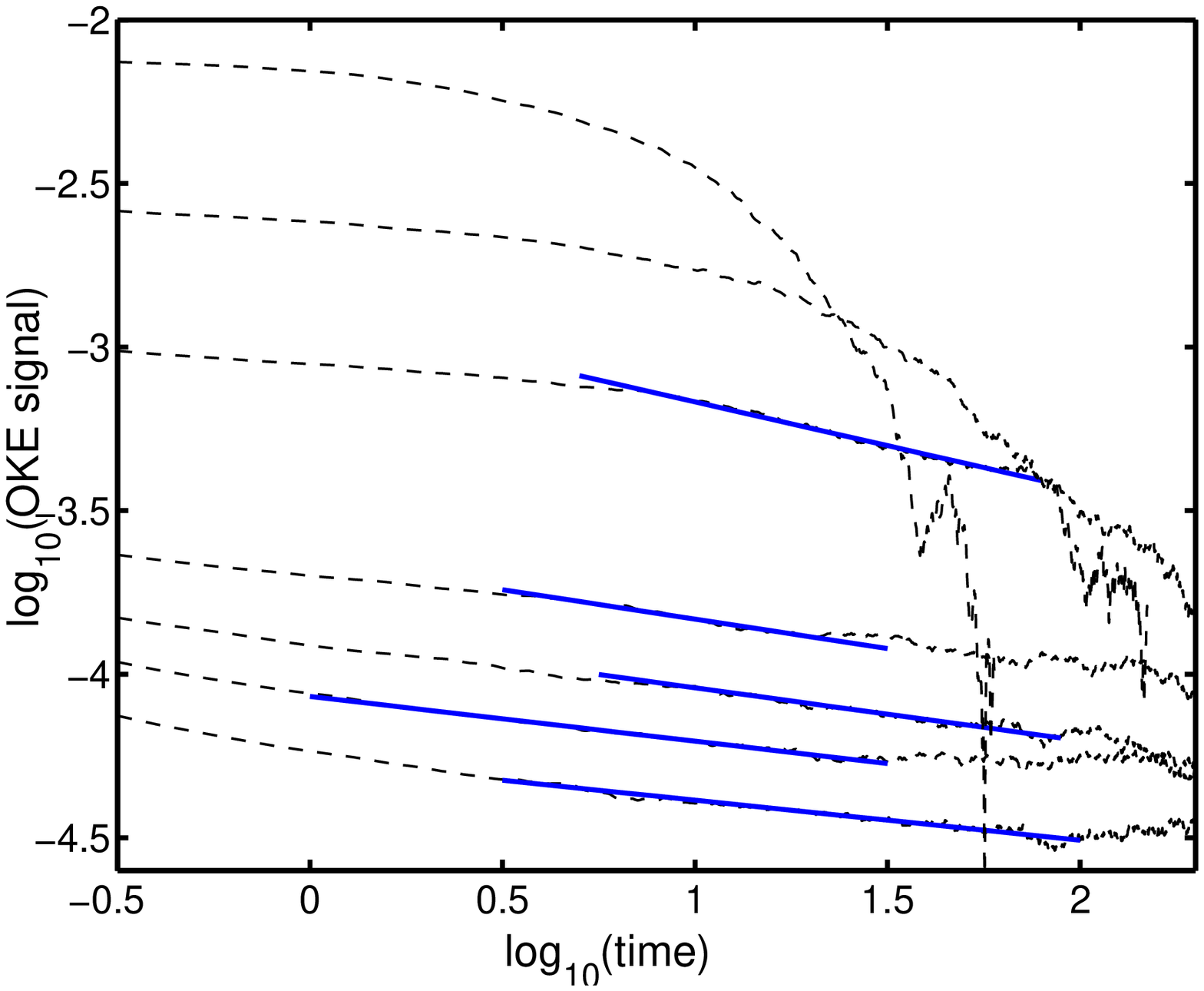,width=7.5cm}
\epsfig{file=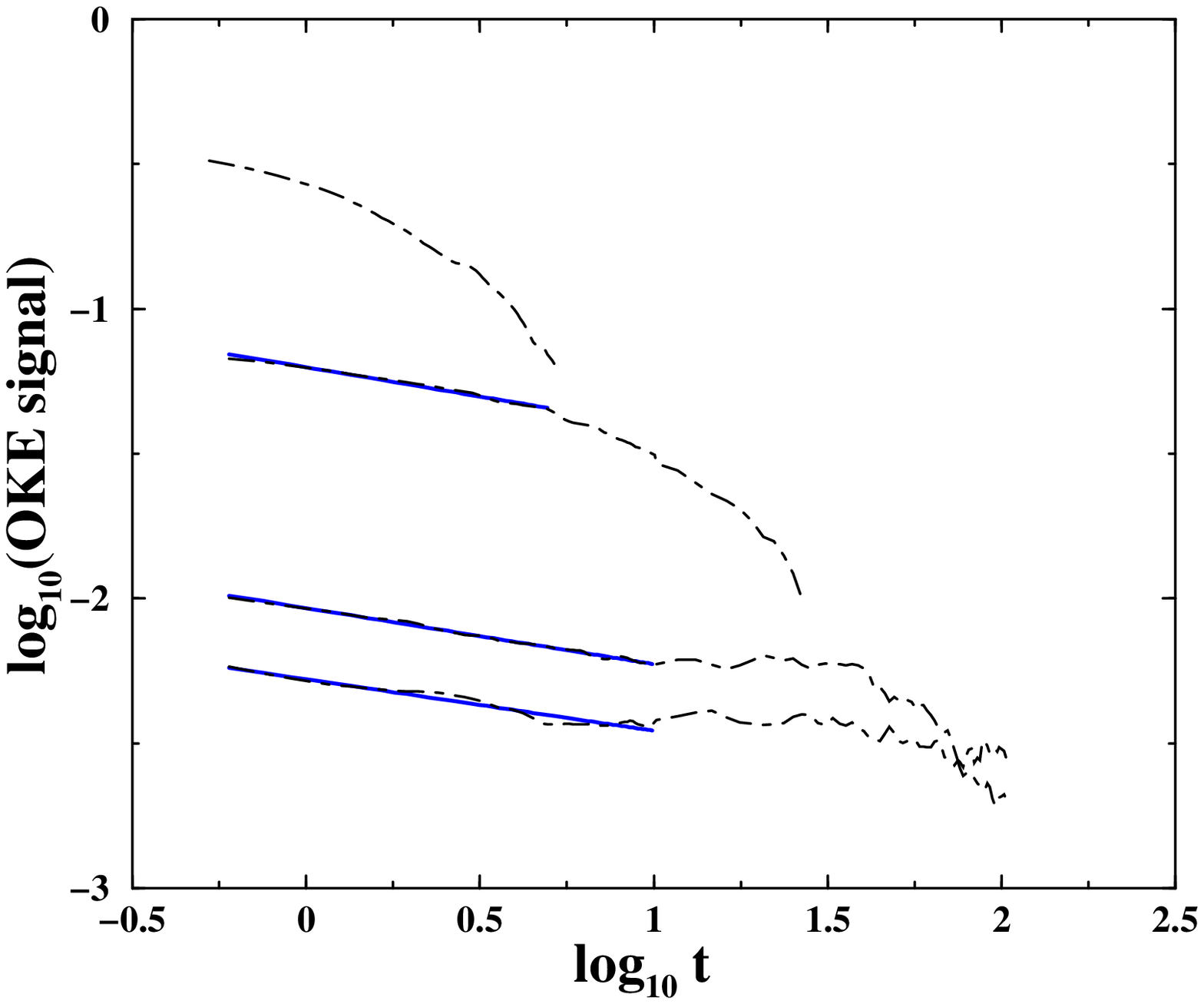,width=7.5cm}
\epsfig{file=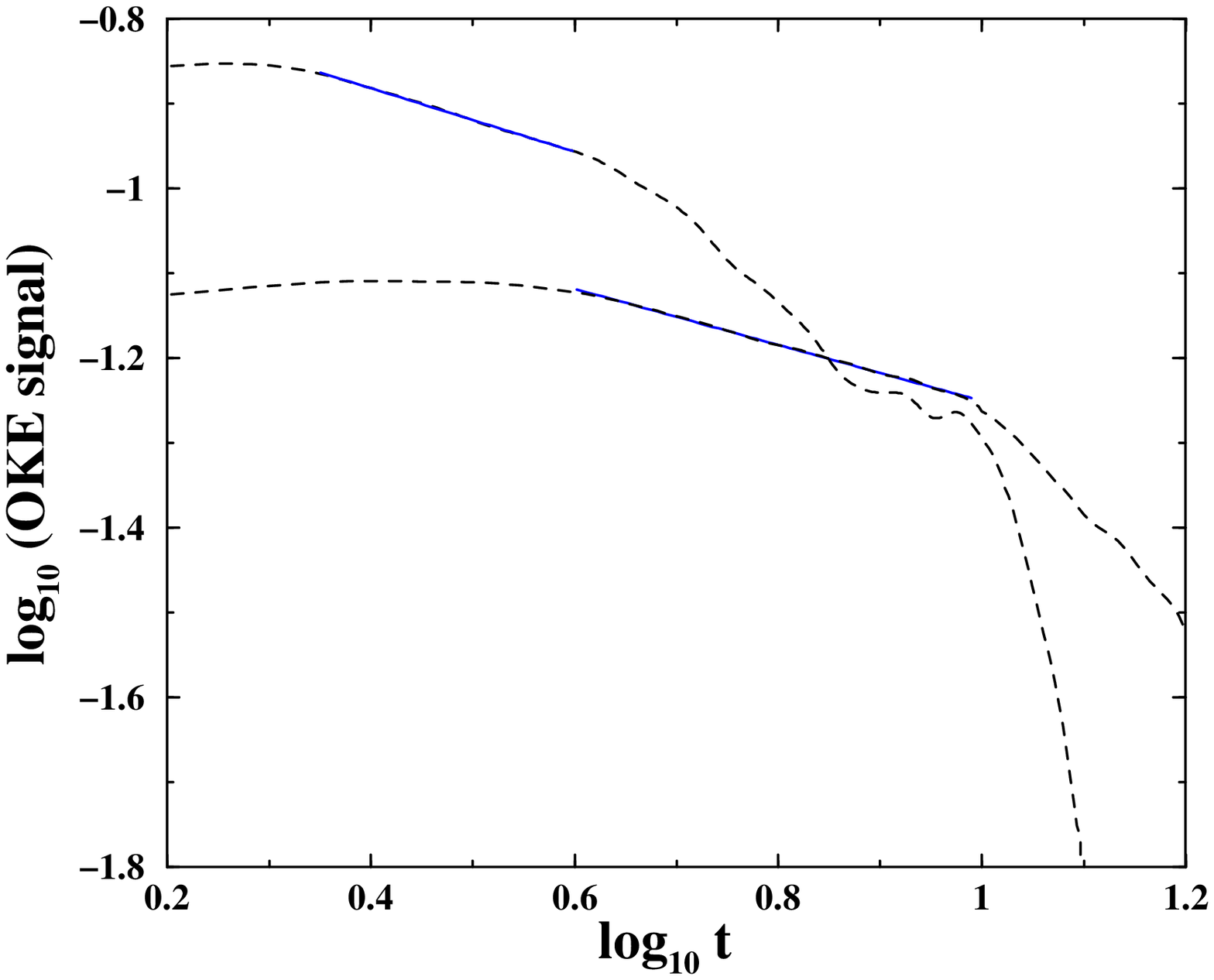,width=7.5cm}
\caption{Time evolution of the collective second rank OTCF at short to 
intermediate times in log-log plots for the (a) calamitic system, (b) discotic 
system, and (c) lattice system. The black lines are simulation data corresponding 
to increasing order parameter from bottom to top. For all the three cases, the 
blue solid lines show the fits to the data over the time regimes, if any, where 
the decay follows a power law: $~ t^{-\alpha}$. The values of the power law 
exponent $\alpha$are given below in the parenthesis. (a) Along an isotherm at 
$T = 1.0$ at several densities: $\rho = 0.285$, $\rho = 0.3$, $\rho = 0.305$ 
($\alpha= 0.26$), $\rho= 0.31$ ($\alpha= 0.18$), $\rho= 0.315$ 
($\alpha= 0.16$), $\rho = 0.32$ ($\alpha = 0.13$), and $\rho = 0.33$ 
($\alpha= 0.12$).  (b) Along an isobar at $P = 25.0$ at several temperatures: 
$T = 2.991$, $T = 2.693$ ($\alpha= 0.208$), $T = 2.646$ ($\alpha= 0.194$), and 
$T = 2.594$ ($\alpha= 0.178$). (c) At two temperatures: $T = 1.176$ 
($\alpha= 0.374$) and $T = 1.149$ ($\alpha= 0.33$).} 
\end{figure}

In experiments, one can probe orientational relaxation through the decay of the 
OKE signal, which is given by the negative of the time derivative of the 
collective second rank OTCF $C_{2}^{c}(t)$ \cite{r7,r8,r9}.  The latter, which is 
defined by
\begin{equation}
C_{2}^{c}(t)={\frac{\langle{\displaystyle \sum_{i} \sum_{j}P_{2}({\bf \hat e}_{i}
(0)\cdot{\bf \hat e}_{j}(t))}\rangle}{\langle{\displaystyle \sum_{i} \sum_{j} 
P_{2}({\bf \hat e}_{i}(0)\cdot {\bf \hat e}_{j}(0))}
\rangle}},
\label{cc2}
\end{equation}
is computationally demanding, particularly at long times.In order to set a direct
link with experimental results, we show the temporal behavior of the OKE signal in
the log-log plots in the respective systems across the I-N phase transition in 
Figs. 2(a), 2(b), and 2(c). {\it The short-to-intermediate-time power law regime}
is evident in the OKE signal for all the three systems. While a power law decay of 
the OKE signal has been recently observed experimentally in calamitic systems 
near the I-N phase boundary and in the nematic phase (8-9,18), our prediction for 
the discotic systems could be tested in experiments.

We now address the origin of this general power law relaxation with
two plausible explanations. The first originates from the observation that such
power laws are well-known in supercooled liquids where an elegant expression is
provided by the mode coupling theory (MCT) \cite{r19,r20}. In the MCT description
\cite{r18}, a non-linear dependence of the memory function (the 
longitudinal viscosity) on the density-density time correlation function and 
the feed-back between the viscosity and the dynamic structure factor lead to the 
power law  relaxation, which is, therefore, purely dynamical in origin.
Since the memory function involved here pertains to orientational motion,
as a first approximation, we may replace it by the rotational friction,
which can be obtained from the evaluation of the torque-torque time correlation 
function (TCF) by using the time dependent density functional theory. In this
approach, the $Y_{lm}$-th component of the wavenumber (k) and orientation 
dependent fluctuating density of the mesogen is the natural slow-variable.
The following expression for the singular part of the memory function can then be 
obtained for axially symmetric mesogens:
\begin{equation}
\Gamma^{Sing}(z) = \frac{3k_{B}T\rho}{8\pi^{3}I}\int_{0}^{\infty} dk k^2 
\displaystyle \sum_{l,m}c_{l,l,m}^{2}(k)S_{l,m}(k,z),
\end{equation}
where $c_{l,l,m}(k)$ is the (l,l,m) component of the spherical harmonic expansion 
of the two particle direct correlation function, $S_{l,m}(k,z)$ is the same for 
the (l,m) component of the orientational dynamic structure factor, and $z$ is the 
Laplace frequency \cite{Gray-book,ACP-Bagchi}. Near the I-N transition dynamic 
structure factor $S_{20}(k,z)$ slows down dramatically due to the rapid and 
diverging growth of $S_{20}(k)$ in the long wavelength limit \cite{Allen-PRL}. It 
can be shown that this leads to a power law behavior of the memory function: 
$\Gamma^{Sing}(z) = A^{\prime} z^{-1/2}$, $A^{\prime}$ being a constant, at short 
to intermediate times. Use of this memory function in the Mori-Zwanzig continued 
fraction representation of the dynamic structure factor gives 
$S_{20}(t) = exp(a^{2}t)erfc(at^{1/2})$, where $a$ is a constant. This has the 
required power law behavior. Thus, here the power law arises due to
diverging correlation length unlike in supercooled liquids where no such 
divergence of correlation length is involved. Near the I-N phase boundary, on the 
other hand, the free energy surface can play a direct role in the short time power
law. Near the I-N transition, an equation of motion for the fluctuating 
orientational order parameter ($\delta S$), a non-conserved variable \cite{r21}, 
can be written as follows:
\begin{equation}
\frac{d \delta S}{dt} = - \int dt \Gamma (t-t^{\prime}) \frac{\delta F}{\delta (\delta S)}(t^{\prime}) + R(t),
\label{ncv}
\end{equation}
where $\Gamma$ is a damping coefficient, $F(\delta S)$ is the Landau-de Gennes 
free energy as a function of the orientational order parameter and $R(t)$ is a 
random velocity term related to $\Gamma$ by the fluctuation-dissipation theorem. 
As the temperature approaches the critical temperature $T_{c}$, the free energy 
surface becomes soft. If one uses the Landau free energy expansion 
$\delta F = A(T)(\delta S)^{2} + B(T) (\delta S)^{3} + C(T) (\delta S)^{4}$, then
it can be shown that Eq. (4) can also give rise to a power law 
decay of $\frac{<\Delta S(0)>}{<\Delta S(t)>}$ at short to intermediate times, 
without the necessity of invoking power law behavior of the memory function. 
Note that both explanations involve diverging second-rank orientational
correlation length $\xi_{2}(T)$, which is in turn attributed to the softening of 
the coefficient $A(T)$ in the Landau-de Gennes theory because 
$A(T) \propto \xi_{2}(T)^{-2}$. However, in Eq. (4) higher order terms in the free
energy expansion are important at short to intermediate times.

\begin{figure}
\epsfig{file=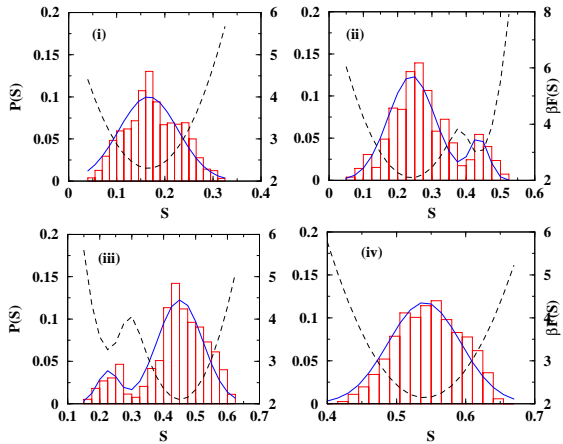,width=7.5cm}
\caption{The distribution $P(S)$ of the orientational order parameter $S$ for a 
256-particle calamitic system at four state points across the I-N transition along
an isochor at the density $\rho= 0.32$: (i) $T = 1.495$ ($<S> = 0.176$), (ii) 
$T = 1.390$ ($<S> = 0.280$), (iii) $T = 1.324$ ($<S> = 0.435$), and (iv) 
$T = 1.191$ ($<S> = 0.547$). The blue solid lines are fits to the histograms
(red solid lines) with a Gaussian function for (i) and (iv) and with a linear 
combination of two Gaussian functions for (ii) and (iii). The black dashed line is
the scaled free energy $F$, where $\beta= 1/(k_{B}T)$, obtained from the fit 
function $P(S)$ through $\beta F(S) = - ln(P(S))$.}
\end{figure}
In order to quantitatively understand the role of order parameter fluctuations, we
have computed the distribution of the fluctuating orientational order parameter 
$S$. In Fig. 3, we show this distribution at four temperatures across the I-N 
phase boundary along an isochor. For the finite size effect, the average 
orientational order parameter has nonzero finite value (of the order of 
$N^{-1/2}$) even in the isotropic phase. While the distribution of $S$ is unimodal in both the isotropic and nematic phases, it becomes
bimodal near the I-N phase boundary with two peaks at $S_{i}$ and $S_{n}$ 
($S_{i} < S_{n}$) corresponding to an isotropic-like and a nematic-like 
configuration, respectively. The dominant peak shifts from $S_{i}$ to $S_{n}$ and 
then the distribution becomes unimodal as the system settles into the nematic 
phase. In the same figure, we have also shown the order parameter dependence of the
free energy obtained from the distribution. Such a bimodal distribution could be 
observed only because of the weakly first order nature of the I-N transition (that
is, the requirement of nucleation is not stringent). {\it Note that the free 
energy barrier between isotropic and nematic phases is only of the order of
$0.01 k_{B}T$ per particle.} Resulting large fluctuations in $S$ can give rise to 
power law decay of $\frac{<\Delta S(0)>}{<\Delta S(t)>}$ as discussed above.

If the power law decay is driven by the growth of the orientational correlation 
length, then its duration is expected to increase with the aspect ratio of the 
mesogens. We have, therefore, carried out a study of rod-like mesogens with aspect
ratio 3.8 and confirmed this expectation.

The emergence of the power law relaxation in the all three systems under 
consideration here near the I-N transition is due to the rapid growth of the 
orientational correlation length, which is also responsible for the universal 
features. Unlike in glassy systems, the power law decay may reflect large 
thermodynamic fluctuations of the order parameters, as shown in Fig. 3. Finally, 
our prediction of power law decay for discotic liquid crystals could be tested in 
experiments.

This work was supported by the grant from DST, India. DC acknowledges UGC, India 
for providing Research Fellowship.


\begin{references}

\bibitem{r1} P. G. de Gennes and J. Prost, {\it The Physics of Liquid Crystals} 
(Clarendon Press, Oxford, 1993).
\bibitem{r2} S. Chandrasekhar, {\it Liquid Crystals} (Cambridge University Press, 
Cambridge, 1992).
\bibitem{r3} W. L. McMillan, Phys. Rev. A {\bf 4}, 1238 (1971).
\bibitem{r4} P. G. de Gennes, Solid State Commun. {\bf 10}, 753 (1972).
\bibitem{r5}  B. I. Halperin, T. C. Lubensky, and S.-K. Ma, Phys. Rev. Lett. 
{\bf 32}, 292 (1974).
\bibitem{r6} A. Yethiraj and J. Bechhoefer, Phys. Rev. Lett. {\bf 84}, 3642 (2000).
\bibitem{r7} Y.-X. Yan and K. A. Nelson, J. Chem. Phys. {\bf 87}, 6240 (1987); 
{\bf 87}, 6257 (1987).
\bibitem{r8} S. D. Gottke, H. Cang, B. Bagchi, and M. D. Fayer, J. Chem. Phys. 
{\bf 116}, 6339 (2002).
\bibitem{r9} J. Li, I. Wang, and M. D. Fayer, J. Phys. Chem. B {\bf 109}, 6514 
(2005).
\bibitem{r10} P. P. Jose and B. Bagchi, J. Chem. Phys. {\bf 120}, 11256 (2004).
\bibitem{r11} J. G. Gay and B. J. Berne, J. Chem. Phys. {\bf 74}, 3316 (1981). 
\bibitem{r12} J. T. Brown, M. P. Allen, E. M. del R\'{i}o, and E. de Miguel, Phys. Rev. E {\bf 57}, 6685 (1998).
\bibitem{r13} M. A. Bates and G. R. Luckhurst, J. Chem. Phys. {\bf 104}, 6696 
(1996). 
\bibitem{r14} P. A. Lebwohl and G. Lasher, Phys. Rev. A {\bf 6}, 426 (1972).
\bibitem{r15} All the quantities reported here for prolate and oblate ellipsoids 
of revolution are given in reduced units defined in terms of the Gay-Berne 
potential parameters $\epsilon_{0}$ and $\sigma_{0}$: length in units of 
$\sigma_{0}$, temperature in units of $\epsilon_{0}/k_{B}$, $k_{B}$ being the 
Boltzmann constant, and time in units of $(\sigma_{0}^{2}m/\epsilon_{0})^{1/2}$, 
$m$ being the mass of the ellipsoids of revolution. We set the mass as well as the
moment of inertia of each of the ellipsoids of revolution equal to unity. In the 
case of the lattice system, temperature is scaled by $\epsilon/k_{B}$. The moment 
of inertia of each particle has been set equal to unity. 
\bibitem{r16} C. Zannoni, {\it Advances in the Computer Simulations of Liquid 
Crystals} (eds. P. Pasini and  C. Zannoni) (Kluwer Academic Publishers, Dordrecht,
2000).  
\bibitem{r17} W. Kob, J. Phys.: Condens. Matter {\bf 11}, R85 (1999).
\bibitem{r18} H. Cang, J. Li, V. N. Novikov, and M. D. Fayer, J. Chem. Phys. 
{\bf 118}, 9303 (2003).
\bibitem{r19} U. Bengtzelins, W. G\"otze, and A. Sj\"{o}lander,  J. Phys. C 
{\bf 17}, 5915 (1984).
\bibitem{r20} W. G\"otze and M. Sperl, Phys. Rev. Lett. {\bf 92}, 105701 (2004).
\bibitem{Gray-book} C. G. Gray and K. E. Gubbins, {\it Theory of Molecular Fluids}
(Clarendon Press, Oxford, 1984), Vol. 1.
\bibitem{ACP-Bagchi} B. Bagchi and A. Chandra, Adv. Chem. Phys. {\bf 80}, 1 
(1991).  
\bibitem{Allen-PRL} M. P. Allen and M. A. Warren, Phys. Rev. Lett. {\bf 78}, 1291
(1997).
\bibitem{r21} P. C. Hohenberg and B. I. Halperin, Rev. Mod. Phys. {\bf 49}, 435 
(1977).

\end{references}
\end{document}